\documentclass[conference]{IEEEtran}
\IEEEoverridecommandlockouts
\usepackage{cite}
\usepackage{amsmath,amssymb,amsfonts}
\usepackage{algorithmic}
\usepackage{graphicx}
\usepackage{textcomp}
\usepackage{xcolor}
\usepackage{hyperref}
\usepackage{float}
\usepackage{multirow}
\def\BibTeX{{\rm B\kern-.05em{\sc i\kern-.025em b}\kern-.08em
    T\kern-.1667em\lower.7ex\hbox{E}\kern-.125emX}}
\begin{document}

\title{A Dual Attention-aided DenseNet-121 for Classification of Glaucoma from Fundus Images}

\author{\IEEEauthorblockN{Soham Chakraborty}
\IEEEauthorblockA{\textit{Dept. of Computer Sc. \& Engg.} \\
\textit{Jadavpur University,} \\
\textit{Kolkata, 700032, India}\\
email: soham.chakraborty@gmail.com}
\and
\IEEEauthorblockN{Ayush Roy}
\IEEEauthorblockA{\textit{Dept. of Electrical Engg.} \\
\textit{Jadavpur University,} \\
\textit{Kolkata, 700032, India}\\
email: aroy80321@gmail.com}
\and
\IEEEauthorblockN{Payel Pramanik}
\IEEEauthorblockA{\textit{Dept. of Computer Sc. \& Engg.} \\
\textit{Jadavpur University,} \\
\textit{Kolkata, 700032, India}\\
email: ppramanik07@gmail.com}
\and
\IEEEauthorblockN{Daria Valenkova}
\IEEEauthorblockA{\textit{Dept. of Automation and Cont. Proc.} \\
\textit{Saint Petersburg Electrotechnical University ”LETI”} \\
\textit{Saint Petersburg, Russia}\\
email: davalenkova@etu.ru}
\and
\IEEEauthorblockN{Ram Sarkar}
\IEEEauthorblockA{\textit{Dept. of Computer Sc. \& Engg.} \\
\textit{Jadavpur University,} \\
\textit{Kolkata, 700032, India}\\
email: ramjucse@gmail.com}
}

\maketitle

\begin{abstract}
Deep learning and computer vision methods are nowadays predominantly used in the field of ophthalmology. In this paper, we present an attention-aided DenseNet-121 for classifying normal and glaucomatous eyes from fundus images. It involves the convolutional block attention module to highlight relevant spatial and channel features extracted by DenseNet-121. The channel recalibration module further enriches the features by utilizing edge information along with the statistical features of the spatial dimension. For the experiments, two standard datasets, namely RIM-ONE and ACRIMA, have been used. Our method has shown superior results than state-of-the-art models. An ablation study has also been conducted to show the effectiveness of each of the components. The code of the proposed work is available at: \url{https://github.com/Soham2004GitHub/DADGC}.
\end{abstract}

\begin{IEEEkeywords}
Glaucoma detection, Fundus image, Deep learning, Dual attention, Channel recalibration 
\end{IEEEkeywords}

 \section{Introduction}\label{intro}
Glaucoma, often referred to as the "silent thief of sight," is a progressive vision loss disease affecting approximately 80 million people worldwide. According to the World Health Organization (WHO), 3.54\% of individuals aged between 40 and 80 are affected by glaucoma, with a higher susceptibility observed in those under 40 compared to those over 80~\cite{mary2016retinal}. Early detection is crucial to initiate preventive measures and mitigate further vision impairment. Diagnosis typically involves examining fundus images for signs such as a pale optic disc (OD), indicative of glaucoma. However, manual assessment by experts is subjective and time-consuming~\cite{sanchez2011evaluation}. To address the increasing prevalence of glaucoma and the need for accessible screening, computer-aided diagnosis (CAD) systems are explored. Nowadays, these systems utilize deep learning techniques to detect glaucoma from fundus images accurately. CAD systems assist in identifying patients requiring immediate examination by an ophthalmologist, thereby reducing the workload on medical professionals while maintaining high sensitivity in diagnosis.

In recent times, Convolutional Neural Networks (CNNs) have become essential in computer vision for tasks like classification and detection~\cite{orlando2017convolutional, pramanik2021deep, pramanik2023deep}. Pre-trained CNN models, originally trained on large datasets, are now utilized for various tasks, often with feature extraction augmented by attention mechanism~\cite{roy2023fourier,roy2024similarity,alirezazadeh2023improving} This enhancement has significantly boosted CNN performance.

\textbf{Contribution:} In this work, we have proposed a transfer learning-based CNN model for glaucoma classification from fundus images. The highlighting points of our work are:

1. DenseNet-121 is used to extract deep features from the input fundus images.

2. We introduce an amalgamation of two attention modules namely, Convolutional Block Attention Module (CBAM) and Channel Recalibration Module (CRM) on the extracted deep features of the DenseNet-121 model. CBAM attention module highlights the relevant spatial and channel features extracted by DenseNet-121. The CRM Module further enriches the features by utilizing edge information along with the statistical features of the spatial dimension.

3. We have evaluated our method on two publicly available datasets namely, RIM-ONE and ACRIMA, and achieved superior results.

\section{Related Work}
In literature of fundus image classification, numerous deep learning-based procedures have been adopted by various researchers. For instance, Sonti et. al~\cite{sonti2023new} introduced KR-NET for retinal fundus glaucoma classification. In this, the region of interest (ROI) is segmented
using variable mode decomposition (VMD) and a 26-layer CNN is designed to classify glaucoma using ACRIMA, RIM-ONE, and Drishti-GS1 datasets. Diaz-Pinto et al.~\cite{diaz2019cnns} utilized CNNs pretrained on ImageNet to automatically assess glaucoma, employing five different models and datasets, including their newly collected dataset called ACRIMA, which is publicly available.
Claro et al.~\cite{claro2019hybrid} created a hybrid space from various feature descriptors with seven CNN architectures, and validated using 10-fold cross-validation (CV) with Random Forest(RF) classifier on various public datasets. Gomez-Valverde et al.~\cite{gomez2019automatic} proposed a method using six different CNN architectures and validated on Drishti-GS1, RIM-ONE datasets using 10-fold CV. Liu et al.~\cite{liu2021small} developed a deep neural network (DNN) model and evaluated on several fundus datasets. An 18-layer CNN model is designed by Elangovan et al.~\cite{elangovan2021glaucoma} for the classification of glaucoma and verified on five public datasets. The authors in~\cite{shyamalee2022cnn} used CNN-based implementation on the ACRIMA dataset and reported a comparative analysis of various CNN models implemented to classify glaucoma. De Sales Carvalho et al.~\cite{de2021automatic} proposed a new method of glaucoma classification using 3-D CNN and verified their results on public datasets with and without segmentation. Elangovan et al.~\cite{elangovan2022convnet} developed a stacking deep ensemble model including 13 pre-trained models and five various approaches for classifying glaucoma images, and validated on publicly available datasets.

Summarizing the literature, a wide range of deep learning approaches has been explored by researchers for glaucoma detection from fundus images. Despite the diversity in methodologies, the primary aim remains constant: to develop reliable models for early glaucoma detection. Our proposed model builds upon this by employing a transfer learning-based CNN approach. We extract deep features using transfer learning and augment them with two attention mechanisms, focusing on relevant spatial and channel features while incorporating edge and statistical features of the spatial dimension.

\section{Methodology}\label{methodology}
\begin{figure}
    \centering
    \includegraphics[width=\linewidth]{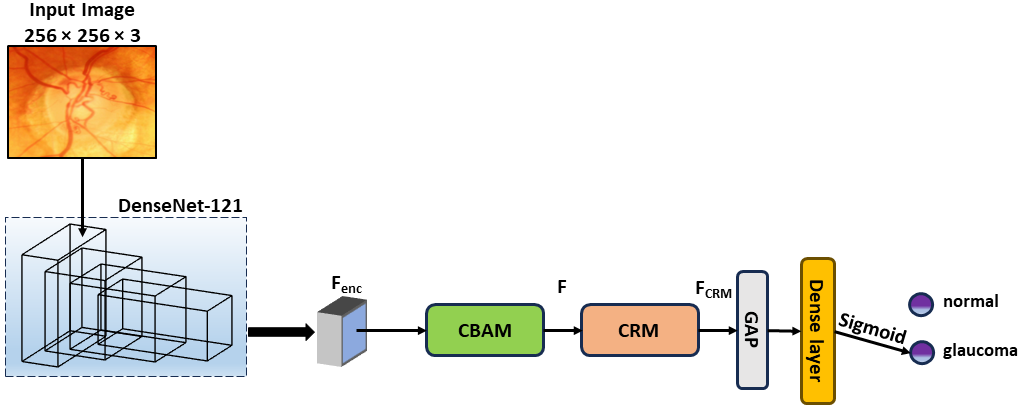}
    \caption{Block diagram of the proposed glaucoma classification model.} 
    \label{fig:architecture}
\end{figure}

To develop a glaucoma classification model, we introduce an amalgamation of the CBAM and CRM attention modules on the encoded features of the DenseNet-121 model, which is the backbone of the entire model. Features generated by the DenseNet-121, denoted as $F_{enc}$, are processed by CBAM and CRM modules successively, and then flattened using a Global Average Pooling (GAP) layer. This flattened feature is subsequently passed to a dense classification layer with Sigmoid activation to predict the class of the inputs. Fig. \ref{fig:architecture} illustrates the proposed model architecture.

\subsection{Convolutional Block Attention Module (CBAM)}

The CBAM attention mechanism \cite{woo2018cbam} enhances feature maps from CNNs by integrating channel-wise and spatial attention. It consists of two main modules: the 1D Channel Attention Module (CAM) and the 2D Spatial Attention Module (SAM). CAM assigns weights to channels, emphasizing those most influential to model performance. The input feature map $F_{enc}$ of CAM undergoes Global Average Pooling (GAP) and Global Max Pooling (GMP) to generate $F_{gap}$ and $F_{gmp}$ respectively. $F_{gap}$ and $F_{gmp}$ are both treated by dense layers separately and then added together. A Sigmoid activation then treats the feature map after addition to generate the channel attention weights. The out of CAM is the channel-wise multiplication of the channel attention weights and $F_{enc}$. The output of CAM enriches the input feature map, which is then processed by SAM. SAM improves feature representation in the spatial dimension by generating a new feature map of the same dimension as the input. It utilizes a convolutional layer followed by dense Layers with rectified linear unit (ReLU) activation to reduce and then restore feature map dimensions. The output of CBAM, enriched across channel and spatial dimensions, focuses specifically on the optic cup and disc in fundus images, aiding in better classification results.

\subsection{Channel Recalibration Module (CRM)}
The CRM adaptively recalibrates feature representations based on channel-wise statistics. It aims to adaptively emphasize informative channels while suppressing less relevant ones. By incorporating both intensity and edge-related information, the CRM enables CNNs to capture fine-grained patterns essential for accurate glaucoma fundus classification.

\begin{figure}
    \centering
    \includegraphics[width=0.74\linewidth]{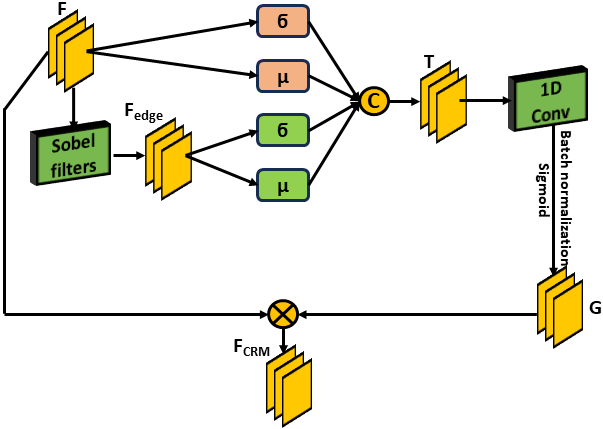}
    \caption{Channel Recalibration Module.} 
    \label{fig:CRM}
\end{figure}

The input feature maps are reshaped to facilitate subsequent operations. $F$ and its corresponding edge map, $F_{edge}$, are reshaped to enable efficient channel-wise computations. The mean $\mu$ and standard deviation $\sigma$ are calculated along the channel axis for both the original and edge feature maps. These statistics capture essential information about spatial feature distribution and gradients, respectively. These $\mu$ and $\sigma$ values are concatenated. This concatenation operation forms a composite tensor $T$ that encapsulates both intensity and edge-related statistics. A 1D convolutional layer (kernel size = 2) is applied to the $T$, followed by batch normalization. This convolutional transformation facilitates the learning of interdependencies between the mean and standard deviation components across channels. The output of the convolutional layer is treated by a Sigmoid activation function, resulting in a gating tensor $G$. This gating tensor modulates the original feature map by assigning importance weights to each channel adaptively. $F$ is recalibrated by performing a hadamard product with $G$, producing $F_{CRM}$. As mentioned earlier, such recalibration enhances informative channels and suppresses irrelevant ones, thereby improving discriminative power in subsequent layers. The block diagram representation of CRM is shown in Fig. \ref{fig:CRM}.

\section{Results}

\subsection{Experimental Setup}
We have conducted research using two standard datasets, ACRIMA \cite{diaz2019cnns} and RIM-ONE \cite{fumero2011rim}, for glaucoma classification.  The ACRIMA dataset comprises 705 images, with 309 healthy and 396 glaucoma images, obtained from a Spanish national project via the IMAGEnet training system. These images, available in .jpg format, range in dimensions from 178 × 178 pixels to 1420 × 1420 pixels. The RIM-ONE dataset, developed in collaboration with three Spanish hospitals, contains 485 images, including 313 healthy and 172 glaucoma images. Similarly centered around the optic disc, these images are available in .png format, with dimensions ranging from 290 × 290 pixels to 1375 × 1654 pixels. For both datasets, we performed a 5-fold cross-validation. Augmenting the training set and standardizing image dimensions to 256 × 256 × 3, we have employed a learning rate of 0.001, the Adam optimizer, and a batch size of 16 on an NVIDIA TESLA P100 GPU. Training the model for 50 epochs using binary cross-entropy loss, we have evaluated its performance using accuracy (Acc), precision (Pre), recall (Rec), and F1-score (F1).

\subsection{Ablation Study}
We have used RIM-ONE as the dataset for the ablation study. For selecting the backbone, we have trained four commonly used CNN models, MobileNetV2, DenseNet-121, ResNet-50, and InceptionV3. Among these, DenseNet-121 gives the best results as shown in Table \ref{ablation1}. Using DenseNet-121 as the backbone, we have performed more experiments to figure out the best architectural configuration. These experiments are: (i) DenseNet-121 + CBAM, (ii) DenseNet-121 + CRM, (iii) DenseNet-121 + CBAM + CRM.

In Table \ref{ablation2}, it is seen that the introduction of CBAM and CRM modules enhances the performance of the baseline, i.e., DenseNet-121. Also, the combined impact of the CBAM and CRM modules produces the best result and further boosts the performance of the baseline. 

\begin{table}[ht!]
    \centering
    \caption{Ablation study for selecting the best baseline model. }    
    \begin{tabular}{lcccc}
        \hline
        \textbf{Model} & \textbf{Acc (\%)} & \textbf{Pre (\%)} & \textbf{Rec (\%)} & \textbf{F1 (\%)}\\      
        \hline
        MobileNetV2 & 84.54 & 84.88 & 83.29 & 83.43\\ 
        ResNet-50 & 87.63 & 86.61 & 86.59 & 86.60\\
        InceptionV3 & 90.52 & 89.59 & 89.22 & 89.40\\
        \textbf{DenseNet-121} & \textbf{90.93} & \textbf{90.73} & \textbf{90.24} & \textbf{90.48}\\ 
        \hline
    \label{ablation1}
    \end{tabular}
\end{table} 

\begin{table}[ht!]
    \centering
    \caption{Ablation study for selecting the best model configuration. }    
    \begin{tabular}{lcccc}
        \hline
        \textbf{Model} & \textbf{Acc (\%)} & \textbf{Pre (\%)} & \textbf{Rec (\%)} & \textbf{F1 (\%)}\\      
        \hline 
        (i) & 90.93 & 90.29 & 88.66 & 89.90\\
        (ii) & 91.96 & 91.98 & 90.80 & 91.13\\
        \textbf{(iii)} & \textbf{93.81} & \textbf{93.40} & \textbf{93.59} & \textbf{93.49}\\
        \hline
    \label{ablation2}
    \end{tabular}
\end{table}

\begin{figure}[ht!]
    \centering
    \includegraphics[width=0.7\linewidth]{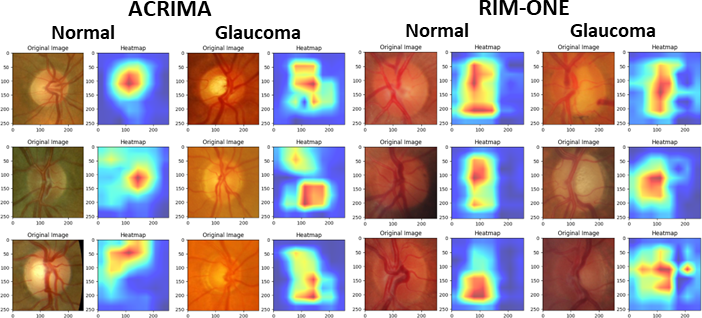}
    \caption{Heatmap of $F_{CRM}$ for normal and glaucoma fundus images.} 
    \label{fig:heat}
\end{figure}

Moreover, the focus of CBAM and CRM can be seen in Fig. \ref{fig:heat}, where the heatmap demonstrates the regions highlighted by these attention modules to streamline the focus of the deep learner.

\subsection{Comparison with SOTA}
It can be seen from Table~\ref{sota_acrima} that the proposed model outperforms the existing models in terms of accuracy and F1 score for ACRIMA. The proposed model outperforms the existing models in terms of accuracy for RIM-ONE but is comparable in terms of F1 score as shown in Table~\ref{sota_rim}.

\begin{table}[htb]
    \centering
    \caption{Performance comparison of the proposed model with SOTA methods on the RIM-ONE dataset.}
    \begin{tabular}{ccc} 
        \hline
        \textbf{Model} & \textbf{Acc(\%)} & \textbf{F1-score(\%)} \\ \hline
        KR-NET, 2023~\cite{sonti2023new} & 90.51 & 89.26\\
        QB-VMD, 2019~\cite{agrawal2019automated} & 86.13 & -\\
        Hybrid PolyNet, 2023~\cite{sangeethaa2023presumptive} & 71.79 & -\\
        Modified CNN, 2021~\cite{elangovan2021glaucoma} & 85.97 & - \\
        EyeNet, 2021~\cite{suguna2021performance} & 89.00 & -\\
        AG-CNN,2019~\cite{li2019attention} & 85.20 & 83.70 \\
        DCNN, 2018~\cite{perdomo2018glaucoma} & 89.40 & -\\
        IEMD,2022~\cite{parashar2022classification} & 93.26 & 92.90 \\
        \textbf{Proposed} & \textbf{93.81} & \textbf{93.49} \\ \hline
    \end{tabular}
    \label{sota_rim}
\end{table}

\begin{table}[htb]
    \centering
    \caption{Performance comparison of the proposed model with SOTA methods on the ACRIMA dataset.}
    \begin{tabular}{ccc} 
        \hline
        \textbf{Model} & \textbf{Acc(\%)} & \textbf{F1-score(\%)} \\ \hline
        KR-NET, 2023~\cite{sonti2023new} & 96.70 & 97.05 \\ 
        Hybrid PolyNet, 2023~\cite{sangeethaa2023presumptive} & 96.21 & - \\
        Modified CNN, 2021~\cite{elangovan2021glaucoma} & 96.64 & 96.89 \\
        DNet-201, 2021~\cite{ovreiu2021deep} & 97.00 & 96.96 \\
        DCGAN, 2022~\cite{singh2022novel} & 93.65 & 94.00\\
        VGG-CapsNet, 2022~\cite{singh2022novel} & 85.85 & 89.00\\
        Hybrid CNN, 2024~\cite{oguz2024cnn} & 92.96 & 93.75 \\
        MAS-aided CNN, 2024~\cite{sonti2024diagnosis} & 97.18 & 97.06 \\
        \textbf{Proposed} & \textbf{98.58} & \textbf{98.55} \\ \hline
    \end{tabular}
    \label{sota_acrima}
\end{table}

\begin{figure}
    \centering
    \includegraphics[width=0.7\linewidth]{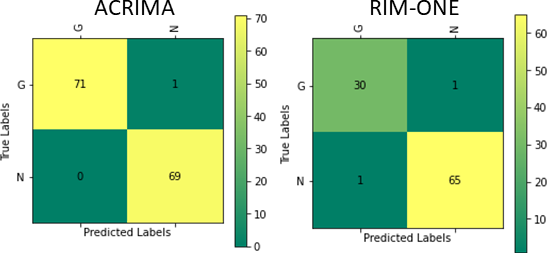}
    \caption{Confusion matrices of the proposed model for ACRIMA and RIM-ONE datasets for the best fold among the 5 folds. 'G' and 'N' indicate 'Glaucoma' and 'Normal' classes respectively.} 
    \label{fig:cm}
\end{figure}

The confusion matrices in Fig.~\ref{fig:cm} demonstrate the effective classification of glaucoma by the proposed model for both ACRIMA and RIM-ONE datasets.

\section{Conclusion and Future scope}
The paper proposes a dual attention-aided DenseNet-121 architecture for classifying glaucoma fundus images. DenseNet-121 acts as a backbone to extract information from the input image. This extracted feature is then spatially and channel-wise enriched using the CBAM and CRM successively. The model demonstrates superior performance than the SOTA methods. Instead of using data augmentation methods, we would like to explore the few-shot learning approaches. Another plan is to use a lightweight backbone to make the model applicable in a resource-constraint environment.  

\bibliographystyle{vancouver}
\bibliography{Bibliography.bib}
\end{document}